\documentclass[nofootinbib]{aastex7}
\usepackage[utf8]{inputenc}
\usepackage{amsmath,units}
\usepackage{booktabs}
\usepackage{makecell} 
\usepackage{color}
\usepackage{bold-extra}
\usepackage[T1]{fontenc}
\usepackage{graphicx}
\usepackage{acronym}   
\usepackage{xspace}
\usepackage{enumitem} 
\usepackage{calc}
\usepackage[caption=false]{subfig}
\usepackage{import}
\usepackage{amsmath}	
\usepackage{bbm} 
\usepackage{mathtools} 
\usepackage{spverbatim}
\newcommand{\optionname}[1]{\textsc{\texttt{#1}}}

\newcommand{\vCOMPAS}{\ensuremath{\rm{v}03.22.01}\xspace} 

\newcommand\COMPAS{{\sc{COMPAS}}\xspace}  

\acrodef{OOP}{object-oriented programming}
\acrodef{ZAMS}{zero-age main sequence}
\acrodef{MS}{main sequence}
\acrodef{HG}{Hertzsprung gap}
\acrodef{HR}{Hertzsprung-Russell}
\acrodef{NS}{neutron star}
\acrodefplural{NS}[NSs]{neutron stars}
\acrodef{BH}{black hole}
\acrodefplural{BH}[BHs]{black holes}
\acrodef{CO}{compact object}
\acrodefplural{CO}[COs]{compact objects}
\acrodef{CHE}{chemically homogeneous evolution}
\acrodef{CHeB}{core helium burning}
\acrodef{AGB}{asymptotic giant branch}
\acrodef{EAGB}{early asymptotic giant branch}
\acrodef{HeMS}{helium main sequence}
\acrodef{HeGB}{helium giant branch}

\acrodef{GW}{gravitational wave}
\acrodef{BBH}{binary black hole}
\acrodefplural{BBH}[BBHs]{binary black holes}
\acrodef{DNS}{double neutron star}
\acrodefplural{DNS}[DNSs]{double neutron stars}
\acrodef{DCO}{double compact object}
\acrodefplural{DCO}[DCOs]{double compact objects}
\acrodef{BH-{}-NS}{black hole-neutron star}
\acrodef{WD}{white dwarf}
\acrodefplural{WD}[WDs]{white dwarfs}
\acrodef{HeWD}{helium white dwarf}
\acrodefplural{HeWD}[HeWDs]{helium white dwarfs}
\acrodef{COWD}{carbon-oxygen white dwarf}
\acrodefplural{COWD}[COWDs]{carbon-oxygen white dwarf}
\acrodef{ONeWD}{oxygen-neon white dwarf}
\acrodefplural{ONeWD}[ONeWD]{oxygen-neon white dwarf}

\acrodef{GRB}{gamma-{}-ray burst}
\acrodef{RLOF}{Roche-lobe overflow}
\acrodef{CE}{common envelope}
\acrodef{SN}{supernova}
\acrodefplural{SN}[SNe]{supernovae}
\acrodef{ECSN}{electron-capture supernova}
\acrodefplural{ECSN}[ECSNe]{electron-capture supernovae}
\acrodef{USSN}{ultra-stripped supernova}
\acrodefplural{USSN}[USSNe]{ultra-stripped supernovae}
\acrodef{CCSN}{core-collapse supernova}
\acrodefplural{CCSN}[CCSNe]{core-collapse supernovae}
\acrodef{PISN}{pair-instability supernova}
\acrodefplural{PISN}[PISNe]{pair-instability supernovae}
\acrodef{LRN}{luminous red nova}
\acrodefplural{LRN}[LRNe]{luminous red novae}
\acrodef{LBV}{luminous blue variable}
\acrodefplural{LBV}[LBVs]{luminous blue variables}
\acrodef{WR}{Wolf-Rayet}

\acrodef{SNR}{signal-to-noise ratio}

\acrodef{COMPAS}{
Compact Object Mergers: Population Astrophysics and Statistics}
\acrodef{BPS}{binary population synthesis} 
\acrodef{SSE}{single star evolution} 
\acrodef{BSE}{binary star evolution} 
\acrodef{LVC}{LIGO-Virgo Collaboration}
\acrodef{LVK}{LIGO-Virgo-KAGRA Collaboration}
\acrodef{LISA}{Laser Interferometer Space Antenna}
\acrodef{IMF}{initial mass function}

\acrodef{GSMF}{galaxy mass function, the number density of galaxies per logarithmic mass bin}
\acrodef{MZR}{mass-metallicity relation}
\acrodef{SFRD}{star formation rate density}

\acrodef{CSV}{comma separated values}
\acrodef{HDF5}{hierarchical data format, version 5}
\acrodef{TSV}{tab separated values}
\acrodef{TXT}{plain text}

\defcitealias{COMPAS:2021}{Paper I}

\begin{document}

\title{Rapid stellar and binary population synthesis with COMPAS: methods paper II}

\shorttitle{COMPAS code paper}
\shortauthors{Team COMPAS}

\author{Team COMPAS}
\email{ilya.mandel@monash.edu}
\affiliation{The public~\COMPAS code is a product of work by the entire \COMPAS collaboration over many years; we therefore kindly request that, in recognition of this team effort, the paper is cited as ``Team \COMPAS: I.~Mandel et al.''}
\author[0000-0002-6134-8946]{Ilya Mandel}
\email{ilya.mandel@monash.edu}
\affiliation{School of Physics and Astronomy, Monash University, Clayton, Victoria 3800, Australia}
\affiliation{OzGrav, Australian Research Council Centre of Excellence for Gravitational Wave Discovery, Australia}
\author{Jeff Riley}
\email{jeff.riley@monash.edu}
\affiliation{School of Physics and Astronomy, Monash University, Clayton, Victoria 3800, Australia}
\affiliation{OzGrav, Australian Research Council Centre of Excellence for Gravitational Wave Discovery, Australia}
\author{Adam Boesky}
\email{apboesky@gmail.com}
\affiliation{Center for Astrophysics $|$ Harvard \& Smithsonian, Cambridge, MA 02138, USA}
\author{Adam Br\v{c}ek}
\email{adam.brcek@monash.edu}
\affiliation{School of Physics and Astronomy, Monash University, Clayton, Victoria 3800, Australia}
\affiliation{OzGrav, Australian Research Council Centre of Excellence for Gravitational Wave Discovery, Australia}
\author{Ryosuke Hirai}
\email{ryosuke.hirai@monash.edu}
\affiliation{Astrophysical Big Bang Laboratory (ABBL), Pioneering Research Institute, RIKEN, Wako, Saitama 351-0198, Japan}
\affiliation{School of Physics and Astronomy, Monash University, Clayton, Victoria 3800, Australia}
\affiliation{OzGrav, Australian Research Council Centre of Excellence for Gravitational Wave Discovery, Australia}
\author{Veome Kapil}
\email{vkapil1@jh.edu}
\affiliation{William H. Miller III Department of Physics and Astronomy, Johns Hopkins University, 3400 N. Charles Street, Baltimore, Maryland, 21218, USA}
\author[0000-0002-6592-2036]{Mike Y. M. Lau}
\email{mike.lau@h-its.org}
\affiliation{Heidelberger Institut f\"{u}r Theoretische Studien, Schloss-Wolfsbrunnenweg 35, 69118 Heidelberg, Germany}
\author{JD Merritt}
\email{jmerritt@uoregon.edu}
\affiliation{Department of Physics, University of Oregon, Eugene, OR 97403, USA}
\author[0000-0002-0125-1472]{Nicol\'as Rodr\'iguez-Segovia}
\email{nj.rsegovia@gmail.com}
\affiliation{School of Science, University of New South Wales, Australian Defence Force Academy, Canberra, ACT 2600, Australia}
\author[0000-0002-4181-8090]{Isobel Romero-Shaw}
\email{ir346@cam.ac.uk}
\affiliation{DAMTP, Centre for Mathematical Sciences, University of Cambridge, Wilberforce Road, Cambridge, CB3 0WA, UK}
\affiliation{Kavli Institute for Cosmology, Madingley Road, Cambridge, CB3 0HA, United Kingdom}
\affiliation{H. H. Wills Physics Laboratory, Tyndall Avenue, Bristol BS8 1TL, UK}
\author[0000-0002-2080-9232]{Yuzhe Song}
\email{yuzhesong@swin.edu.au}
\affiliation{Centre for Astrophysics and Supercomputing, Swinburne University of Technology, Hawthorn, VIC 3122, Australia}
\affiliation{OzGrav, Australian Research Council Centre of Excellence for Gravitational Wave Discovery, Australia}
\author[0000-0002-6100-537X]{Simon Stevenson}
\email{spstevenson@swin.edu.au}
\affiliation{Centre for Astrophysics and Supercomputing, Swinburne University of Technology, Hawthorn, VIC 3122, Australia}
\affiliation{OzGrav, Australian Research Council Centre of Excellence for Gravitational Wave Discovery, Australia}
\author[0000-0002-4146-1132]{Avi Vajpeyi}
\email{avi.vajpeyi@auckland.ac.nz}
\affiliation{Department of Statistics, University of Auckland, 38 Princes St, Auckland, New Zealand}
\author[0000-0001-5484-4987]{L.~A.~C.~van~Son}
\email{lvanson@flatironinstitute.org}
\affiliation{Center for Computational Astrophysics, Flatiron Institute, 162 Fifth Avenue, New York, NY 10010, USA}
\affiliation{Department of Astrophysical Sciences, Princeton University, 4 Ivy Lane, Princeton, NJ 08544, USA}
\author[0000-0003-1817-3586]{Alejandro Vigna-G\'{o}mez}
\email{avigna@mpa-garching.mpg.de}
\affiliation{Max-Planck-Institut f\"ur Astrophysik, Karl-Schwarzschild-Str. 1, 85748 Garching, Germany}
\author{Reinhold Willcox}
\email{reinhold.willcox@kuleuven.be}
\affiliation{Institute of Astronomy, KU Leuven, Celestijnenlaan 200D, 3001, Leuven, Belgium}
\affiliation{Leuven Gravity Institute, KU Leuven, Celestijnenlaan 200D, box 2415, Leuven, Belgium}
    
\date{\today}


\begin{abstract}
The \href{https://compas.science}{\COMPAS} public rapid binary population synthesis code has undergone a number of key improvements since the original \COMPAS methods paper \citep{COMPAS:2021} was published.  These include more sophisticated and robust treatments of binary interactions: mass transfer physics, common-envelope events, tides and gravitational-wave radiation reaction; and updated prescriptions for stellar evolution, winds and supernovae.  The code structure and outputs have also been updated, with a focus on improving resolution without sacrificing computational speed.  This paper describes the substantive changes in the code between the previous methods paper and \COMPAS \vCOMPAS.
\end{abstract}

\section{Introduction}

The \COMPAS (Compact Object Mergers: Population Astrophysics and Statistics) rapid binary population synthesis toolkit was initially developed to explore the formation of merging compact object binaries emitting gravitational waves through isolated binary evolution \citep{Stevenson:2017}.  It has since been significantly extended and used to explore gravitational-wave astronomy, Galactic double neutron stars \citep{VignaGomez:2018}, supernova varieties, X-ray binaries, luminous red novae and common envelopes, stellar mergers and cluster populations, and other topics in stellar and binary evolution. 

The methodology and implementation of the code is described in detail by \citet{COMPAS:2021} (methods Paper I), which addresses the development of the code through version 02.21.00.  A slightly later version of the code, 02.27.00, was peer reviewed and briefly summarised by \citet{COMPAS:2022}.  The basic structure of the code has since remained the same and is adequately presented in \citet{COMPAS:2021}.   Here, we only describe the significant changes in \COMPAS between versions 02.21.00 and \vCOMPAS: the key new capabilities and options which have allowed for increasingly sophisticated treatments of binary and stellar evolution.  

The present methods Paper II is limited to substantive changes and does not describe minor modifications or defect repairs; a full record of changes to the main \texttt{dev} branch can be found in the public code repository, \url{https://github.com/TeamCOMPAS/COMPAS}, particularly in \texttt{changelog.h}, while the online documentation \url{https://compas.readthedocs.io/en/latest/} contains detailed descriptions of inputs and outputs and a list of key changes.  

\begin{figure}[t]
\centering
\includegraphics[width=0.5\linewidth]{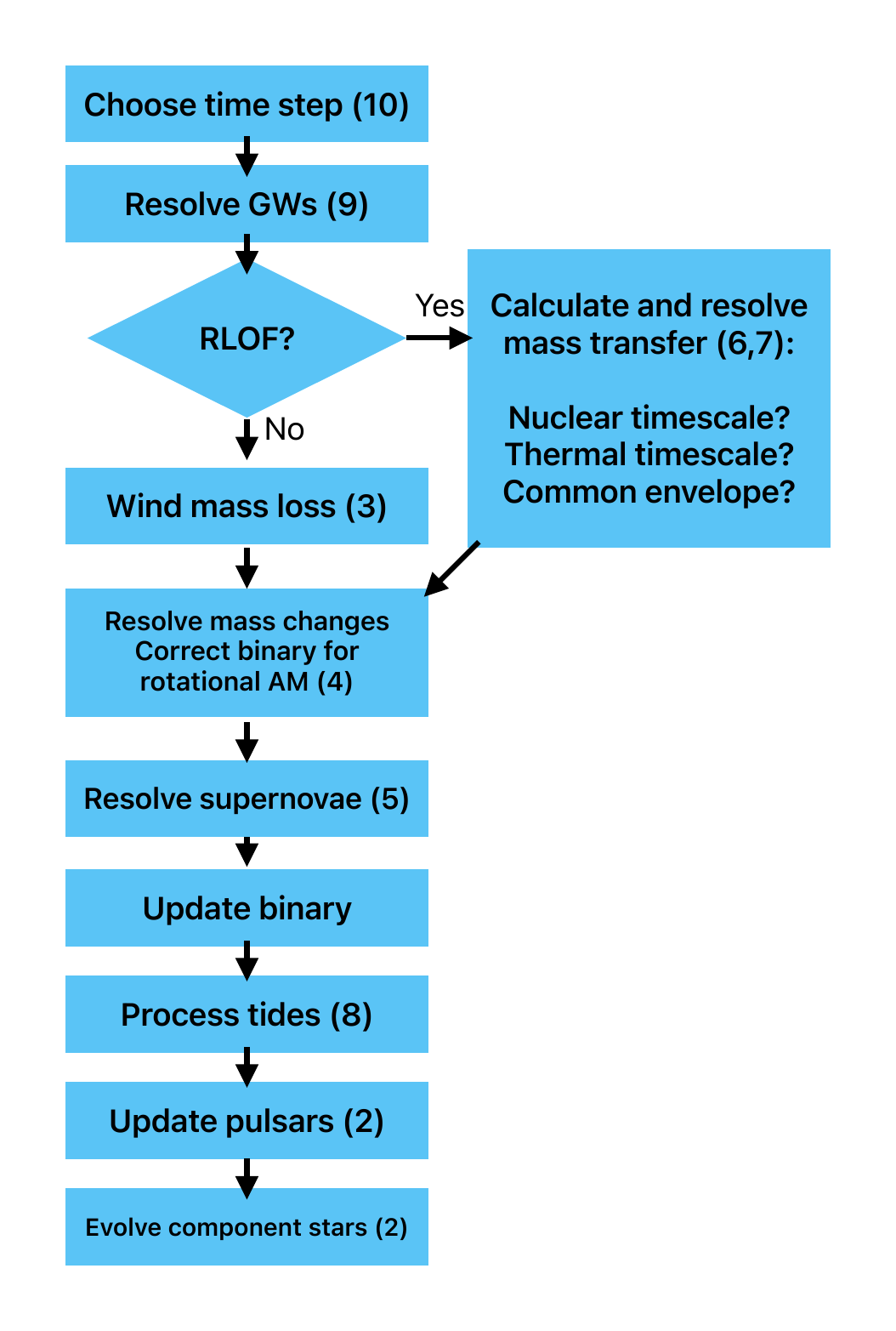}
\caption{Simplified flowchart illustrating one step of COMPAS binary evolution \citepalias[cf.~Figure 4 of ][]{COMPAS:2021}. 
It does not include checks for whether to terminate evolution, e.g., on touching/merging stars, double compact objection formation, or unbound binaries (unless relevant flags allow for continuing evolution).  The numbers in parentheses refer to the sections of this paper that describe the relevant changes.}\label{flowchart}
\end{figure}

For convenience, we divide our description into several key themes: stellar evolution (section \ref{sec:SE}), winds (section \ref{sec:winds}), stellar rotation (section \ref{sec:rot}),  supernovae (section \ref{sec:SN}), mass transfer (section \ref{sec:MT}), common envelopes (section \ref{sec:CE}), tides (section \ref{sec:tides}), gravitational waves (section \ref{sec:GW}), and general code structure improvements (section \ref{sec:code}).  
Figure \ref{flowchart} illustrates the sequence of these calculations in a single time step of binary evolution as modelled by \COMPAS.

\section{Stellar evolution}\label{sec:SE}

\subsection{Convective envelopes}

We included new fits to the masses and binding energies of the convective outer portions of stellar envelopes which are critical for the treatment of tides (see section \ref{sec:tides}) and the 2-stage common envelope formalism (see section \ref{sec:CE}).  The stellar evolution models of \citet{Hurley:2000} used in \COMPAS only contain fits for the total envelope mass, so we previously assumed that the entire envelope abruptly becomes convective at some point in the star's evolution, such as the transition from the Hertzsprung gap to the giant branch or the time when the envelope reaches a given temperature.   We now use convective-envelope mass and binding energy fits provided in section 3.3 of \citet{Picker:2024} as a function of stellar mass, metallicity, and effective temperature.  The fit for the onset temperature of the convective envelope follows Eq.~(6) of \citet{Mandel:2024CE} rather than Eq.~(6) of \citet{Picker:2024} to avoid issues caused by differences between temperatures in MESA models (used in \citealt{Picker:2024} fits) and the \citet{Hurley:2000} tracks used in \COMPAS.

For the tides treatment in particular, we also need to estimate the radial extent of the outer convective zone.  Since rapid models for evaluating this are not available, we assume, inspired by \citet{Hurley:2000,Hurley:2002}, that the radial extent of the convective envelope for Hertzsprung gap and giant stars is given by
\begin{equation}
R_\textrm{conv.~env.} = \sqrt{\frac{M_\textrm{conv.~env.}}{M_\textrm{conv.~env., max}}} \left(R_\mathrm{total}-R_\mathrm{core} \right),
\end{equation}
where $M_\textrm{conv.~env.}$ and $M_\textrm{conv.~env., max}$ are the current and maximal mass of the convective envelope (Eqs.~(7) \& (9), respectively, of \citealt{Picker:2024}) while $R_\mathrm{total}$ and $R_\mathrm{core}$ are the total stellar radius and core radius, respectively.

We now track core and surface hydrogen and helium abundances for all stars with a simplistic model (see, e.g., section \ref{sec:MT}).

\subsection{Pulsation}

To account for the possible emergence of pulsation-driven `superwinds' in red supergiant stars with high luminosity-to-mass ratios \citep{Heger:1997,YoonCantiello:2010}, we added the option to eject the convective envelope of giant stars through dynamical pulsations.  This optional behaviour is turned on with the \texttt{-{}-expel-convective-envelope-above-luminosity-threshold} option and is active if the ratio of luminosity (in $\log_{10} L / L_\odot$ units) to the mass (in $\log_{10} M/M_\odot$ units) exceeds a user-defined threshold \texttt{-{}-luminosity-to-mass-threshold}, set to 4.2 by default following the unpublished work of Matthew Clayton \citep[][section 5.2.3]{Clayton:thesis} and Philipp Podsiadlowski.

\subsection{Neutron stars}

We updated the treatment of the evolution of neutron star spins and magnetic fields.  Isolated pulsars spin down and their fields decay following the treatment of \citet{Oslowski:2011, Song:2024}.  The spins and magnetic fields of accreting neutron stars are evolved by solving differential equations (5) of \citet{Song:2024} and (12) of \citet{Debatri:2020}, respectively, with an optional distinct treatment for a neutron star accreting during a common envelope, set via new option \texttt{-{}-neutron-star-accretion-in-ce}.  The initial magnetic field and spin distributions of newly born neutron stars can be chosen to be flat-in-log, uniform, or log-normal for magnetic fields, and uniform or normal for spin periods; if default log-normal magnetic field and normal spin period distributions are used, their means and standard deviations can be set via new command-line options.  Single neutron stars can now continue to be evolved after formation in the SSE (single stellar evolution) mode of \COMPAS if \texttt{-{}-evolve-pulsars} option is set.

\subsection{Chemically homogeneous evolution}

The lifetimes of chemically homogeneously evolving (CHE) stars can be optionally extended relatively to normal main sequence stars by a factor of
\begin{equation}
\log \frac{\tau_\mathrm{CHE}}{\tau_\mathrm{MS}} = -0.15929 + 1.0500 \log{\frac{M}{M_\odot}} - 0.82336 \left(\log{\frac{M}{M_\odot}}\right)^2 + 0.17772 \left(\log{\frac{M}{M_\odot}}\right)^3
\end{equation}
based on a fit to the models of \citet{Szecsi:2022}, where $M$ is the stellar mass.  Here and elsewhere, all logarithms are base 10.  Meanwhile, luminosities of CHE stars are increased (but never decreased in \COMPAS) relative to normal main sequence \citet{Hurley:2000} models by a factor of
\begin{equation}
\frac{\log L_\mathrm{CHE}}{\log L_\mathrm{MS}} = 1 + \left(0.8261 - \log{\frac{M}{M_\odot}} + 0.58763 \left(\log{\frac{M}{M_\odot}}\right)^2 - 0.10236 \left(\log{\frac{M}{M_\odot}}\right)^3\right) \tau_\mathrm{frac}^2,
\end{equation}
where $\tau_\mathrm{frac}$ is the fractional main sequence age, based on a fit to the models of \citet{Szecsi:2022}.  Both adjustments require the option \texttt{-{}-enhance-CHE-lifetimes-luminosities} (on by default) and have the effect of increasing the total amount of mass lost during CHE. See also additional CHE related adjustments in section \ref{sec:winds}.


\section{Winds}\label{sec:winds}

A new suite of wind mass loss rate models relevant for massive stars has been added to \COMPAS.  These are described in detail by Merritt et al.~(in prep.), so we provide only a brief summary here.

For very massive ($M>100 M_\odot$) main sequence star winds, we implemented mass loss rate prescriptions by \optionname{VINK2011} \citep{Vink:2011}, \optionname{BESTENLEHNER2020} \citep{Bestenlehner:2020}, and \optionname{SABHAHIT2023} \citep{Sabhahit:2023}, with the latter as the default.  The user can specify the chosen prescription with the \texttt{-{}-VMS-mass-loss-prescription} option.

For Wolf-Rayet star winds, \COMPAS previously adopted the prescriptions by \optionname{BELCZYNSKI2010} \citep{Belczynski:2009}.   We have now implemented additional options: \optionname{SANDERVINK2023} (which uses the greater mass-loss rate between \citealt{Vink:2017} or \citealt{SanderVink:2020} as corrected by \citealt{Sander:2023}), \optionname{SHENAR2019} (the greater rate between \citealt{Shenar:2019} and \citealt{Vink:2017}).  We apply these to naked helium stars, with the specific prescription chosen by the user via the \texttt{-{}-WR-mass-loss-prescription} option; \optionname{SANDERVINK2023} is the default choice.

Giant stars with a hydrogen-rich envelope and an effective temperature below 8,000 K and zero-age main sequence mass above $8 M_\odot$ lose mass at a rate given by one of the newly implemented red supergiant wind prescriptions: \optionname{VINKSABHAHIT2023} \citep{VinkSabhahit:2023}, \optionname{BEASOR2020} \citep{Beasor:2020}, \optionname{DECIN2023} \citep{Decin:2023}, \optionname{YANG2023} \citep{Yang:2023}, \optionname{KEE2021} \citep{Kee:2021}, or the older \optionname{NJ90} \citep{NieuwenhuijzenDeJager:1990} prescription.  The choice is specified by the \texttt{-{}-RSG-mass-loss-prescription} option, with \textrm{\optionname{DECIN2023}} as the default.

Stars with an effective temperature above 8,000 K that do not fall into one of the classes listed above lose mass in winds at a rate set by one of the OB mass loss rate prescriptions: \optionname{VINK2001} \citep{Vink:2001}, \optionname{BJORKLUND2022} \citep{Bjorklund:2022}, \optionname{KRTICKA2018} \citep{Krticka:2018}, \optionname{VINK2021} \citep{Vink:2021}, with the latter serving as the default unless over-ridden with the \texttt{-{}-OB-mass-loss-prescription} option.  In the \optionname{VINK2001} prescription, the terminal velocity is scaled by metallicity to the power set by the new \texttt{-{}-scale-terminal-wind-velocity-with-metallicity-power} option (default setting of $0$).

Finally, stars that exceed the Humphreys-Davidson limit \citep[][]{Humphreys:1979ApJ} are assumed to become luminous blue variables (LBVs) and experience eruptive mass loss following the prescription from Paper I.   Wind mass loss is capped at a maximum rate of $0.1 M_\odot$ yr$^{-1}$ regardless of the treatment.

Mass loss for CHE stars benefitted from two additional improvements.  We implemented the \citet{Langer:1998} fit for enhanced mass loss rates from stars rotating at a significant fraction of the break-up velocity, enabled with the \texttt{-{}-enable-rotationally-enhanced-mass-loss} option.  We take a weighted average of OB or very-massive-star winds and Wolf-Rayet winds with a weight based on the current helium fraction following the fit of \citet{Yoon:2006} if the \texttt{-{}-scale-CHE-mass-loss-with-surface-helium-abundance} option is used (on by default).


\section{Rotation}\label{sec:rot}

Stellar rotation was not carefully tracked in earlier versions of \COMPAS.  We now track a star's angular momentum throughout its evolution while assuming rigid body rotation, corresponding to very efficient angular momentum transport.  In the absence of mass change or tides (see section \ref{sec:tides}), this angular momentum is conserved, although the angular frequency $\Omega$ will change as the moment of inertia, which is calculated according to \citet{Hurley:2000}, evolves.

Mass loss through winds or mass transfer carries away the specific angular momentum of the outermost shell of the star, $l = - 2/3 R_*^2 \Omega$, where $R_*$ is the stellar radius before mass loss.  However, stars that lose their entire envelope in one time step (thermal or dynamical timescale mass transfer from an evolved donor) are assumed to do so sufficiently quickly that angular momentum transport is inefficient, so their remaining core continues to rotate with the pre-mass transfer frequency.

Mass gain through mass transfer is assumed to bring in the specific angular momentum of a disk extending down to the stellar surface, $l = \sqrt{G M_* R_*}$, where $M_*$ is the accretor's mass.  We consider several possibilities for the behaviour of stars that may be spun up to super-critical rotation by accretion, with the choice determined by the \texttt{-{}-response-to-spin-up} option.  With the \optionname{NO\_LIMIT} choice, critical rotation $\Omega_c =\sqrt{G M_* / R_*^3}$ is ignored, i.e., the accretor is allowed to spin up to $\Omega > \Omega_c$.  If the \optionname{KEPLERIAN\_LIMIT} option is chosen, mass transfer becomes non-conservative once the accretor (approximately) reaches super-critical rotation; excess mass that is not accreted is assumed to leave the binary with the specific angular momentum of the accretor.  This is approximate because stellar parameters, particularly stellar radius and hence the critical rotation frequency, are updated only after the mass transfer phase.  The default choice, \optionname{TRANSFER\_TO\_ORBIT}, assumes that efficient angular momentum coupling between the accretion stream and the accretor \citep[e.g.,][]{PophamNarayan:1991} allows the accretor to continue gaining mass, with mass transfer efficiency determined without accounting for stellar rotation, while limiting the accretor's rotation frequency to the critical value \citep{Paczynski:1991}; the excess angular momentum is deposited into the orbit.

Stable mass transfer conserves the total angular momentum of the system and ejected material, if any.  In practice, we solve for the separation after mass transfer without accounting for the rotational angular momentum as described in Paper I, then adjust the orbital separation by ensuring that total angular momentum is conserved after accounting for the lost or gained stellar rotational angular momentum as described above.  One practical consequence of this operator-splitting approach is that, although stable mass transfer in \COMPAS strips the donor until it just fills its Roche lobe (see section \ref{sec:MT}), this holds only approximately following the subsequent adjustment to the orbit.

Chemically homogeneous evolution generally proceeds as described in Paper I.  If the orbital frequency at initialisation would exceed the threshold for CHE \citep{Riley:2020}, the star's rotational frequency is set equal to the orbital frequency under the assumption that tides would efficiently spin up the star regardless of the tides model unless the rotational frequency has been explicitly specified by the user; this initialisation step alone does not conserve angular momentum.  Subsequently, the rotation rate of CHE stars evolves as usual under the influence of winds and tides, unless the \optionname{NONE} tides prescription is used (see section \ref{sec:tides}), in which case CHE stars are kept in co-rotation with the binary while conserving total angular momentum.


\section{Supernovae}\label{sec:SN}

A number of improvements to stellar evolution relate specifically to supernova explosions and core collapse, so we list them in a separate section.

Based on the observed pulsar velocity distribution \citet{Willcox:2021} proposed that supernovae imparting very low natal kicks, which we associate with electron capture supernovae, only happen in significant numbers to progenitors that have been stripped of their hydrogen envelopes, in addition to being in the correct core mass range.  We implemented this restriction in \COMPAS as a default.  Users who do wish to allow hydrogen-rich progenitors to experience electron-capture supernovae can do so with the \texttt{-{}-allow-non-stripped-ECSN} option.

We changed the default remnant mass and natal kick distribution to follow the stochastic recipes introduced by \citet{MandelMueller:2020}, retaining all of the previously existing options.  The default neutron star natal kick multiplier was changed to 520 km s$^{-1}$ as calibrated against single-pulsar velocity observations \citep{Kapil:2022}, but can be adjusted with the \texttt{-{}-muller-mandel-kick-multiplier-NS} option (there is a similar option for black hole natal kicks, \texttt{-{}-muller-mandel-kick-multiplier-BH}, where the default parameter value is 200 km s$^{-1}$).  The spread in the kick distribution can be separately adjusted for neutron star and black hole natal kicks with the \texttt{-{}-muller-mandel-sigma-kick-NS} and \texttt{-{}-muller-mandel-sigma-kick-BH} options, respectively, both at $0.3$ by default.

Following the observation of \citet{DisbergMandel:2025} that the \citet{Hobbs:2005} fit to single pulsar velocities misses a Jacobian in the calculation, we corrected the distribution width when the \optionname{MAXWELLIAN} prescription is used for the neutron star natal kick from 265 km s$^{-1}$ to 217 km s$^{-1}$.  We also implemented the \optionname{LOGNORMAL} neutron star natal kick distribution proposed by \citet{DisbergMandel:2025}, Eq.~(5), selectable via the \texttt{-{}-kick-magnitude-distribution} option.

We introduced a new remnant mass prescription for core-collapse supernovae, \optionname{FRYER2022}, which follows \citet{Fryer:2022}.  This prescription has two new options: \texttt{-{}-fryer-22-fmix} and \texttt{-{}-fryer-22-mcrit}, which set the values of $f_\mathrm{mix}$ (default value 0.5) and $M_\mathrm{crit}$ (default value $5.75 M_\odot$) in Eq.~(5) of \citet{Fryer:2022}, respectively.

We also implemented the \optionname{MALTSEV2024} prescription for supernova remnant masses, which follows \citet{Maltsev:2025}.  As the \optionname{FRYER2022} prescription, this is chosen with the \texttt{-{}-remnant-mass-prescription} option.  This prescription has two carbon-oxygen core mass ranges over which stars experience complete collapse, separated by a window of successful explosions leading to the formation of neutron stars or partial fallback black holes.  The mass ranges for these outcomes are sensitive to metallicity and mass transfer history \citep{Maltsev:2025}.

\citet{Hendriks:2023} provide a prescription for the remnant masses of pulsational pair instability supernovae, which has been implemented as a new choice \optionname{HENDRIKS} for the \texttt{-{}-pulsational-pair-instability-prescription} option.  It comes with a new optional parameter (default value 0) which sets $\Delta M_\mathrm{PPI,\ CO\ shift}$ in Eq.~(6) of \citet{Hendriks:2023} via the \texttt{-{}-PPI-CO-Core-Shift-Hendriks} option.

We implemented ``rocket kicks'' for neutron stars that continue to accelerate after the natal kick it receives in a supernova, following \citet{Hirai:2024}.  These kicks are only enabled if non-zero rocket kick magnitudes are set for one or both stars via the \texttt{-{}-rocket-kick-magnitude-1} and \texttt{-{}-rocket-kick-magnitude-2} options; additional new options control the directions of rocket kicks.

Since we anticipate that users are likely to want to continue the evolution of binaries that were unbound by supernova kicks in order to explore the fate of the second companion, we now evolve such binaries by default until a second compact object is formed or the evolution duration is exceeded.  Users not interested in continuing the evolution of unbound binaries should set the \texttt{-{}-evolve-unbound-systems} option to false.

Finally, as part of the improved functionality for accretion onto white dwarfs (see section \ref{sec:MT}), we added or improved the treatment of helium shell detonation, accretion induced collapse, and type Ia supernovae.  These changes are described in more detail below.


\section{Mass transfer}\label{sec:MT}

Mass transfer treatment was significantly updated in \COMPAS since Paper I.  We describe updates to dynamically stable mass transfer in this section and split off updates to the treatment of common-envelope episodes to section \ref{sec:CE}.

We now distinguish between nuclear timescale and thermal timescale mass transfer.  Mass transfer can proceed on a nuclear timescale if the thermal-equilibrium value of $\zeta_* \equiv d \ln R_* / d \ln M_*$ exceeds the rate of response of the Roche lobe to mass transfer, $\zeta_\mathrm{RL} \equiv d \ln R_\mathrm{RL} / d \ln M$, where $R_\mathrm{RL}$ is the donor's Roche lobe radius.  Nuclear timescale mass transfer is in principle allowed for both main sequence and evolved donors.  The actual nuclear timescale mass transfer rate is determined by the requirement that the donor must fit into its Roche lobe at the end of the evolutionary timestep, so is set to the ratio of the required donor mass change to the timestep duration.

When the mass transfer is stable but non-conservative, the specific angular momentum of mass lost from the binary can be fixed to a value between the specific angular momentum of the accretor and the value at the L2 Lagrange point \citep{Willcox:2023}.  With this prescription, set with the \optionname{MACLEOD\_LINEAR} argument to the \texttt{-{}-mass-transfer-angular-momentum-loss-prescription} option, the specific angular momentum of the ejected material in units of the binary's specific angular momentum is fixed to
\begin{equation}
\gamma = \left(\frac{1}{1+q} (1-f_\mathrm{Macleod}) + 2^{1/4} f_\mathrm{Macleod}\right)^2 \frac{(1+q)^2}{q},
\end{equation}
where $q \equiv M_\mathrm{accretor} / M_\mathrm{donor}$ is the mass ratio and $f_\mathrm{Macleod}$ can be separately set for degenerate and non-degenerate accretors with the \texttt{-{}-mass-transfer-jloss-macleod-linear-fraction-degen} and \texttt{-{}-mass-transfer-jloss-macleod-linear-fraction-non-degen} options, respectively.  The default value for both of variants of $f_\mathrm{Macleod}$  is $0.5$, where $0$ corresponds to isotropic re-emission from the accretor and $1$ corresponds to L2 mass loss.

The \citet{Hurley:2002} prescriptions for stars losing mass on the main sequence previously used in \COMPAS switch the mass-losing star to the stellar track of a star with the newly reduced mass, since \citet{Hurley:2000} main sequence models only exist for stars without mass loss.  However, main-sequence donors evolve quite differently from non-mass-losing stars, retaining a larger convective core than stars of the new mass \citep{Shikauchi:2024}, which ultimately leads to higher remnant masses for mass-losing stars.  We added a new option \texttt{-{}-main-sequence-core-mass-prescription} to force stars to retain a greater core mass following main sequence mass loss.  In the \optionname{MANDEL} variant, which is the current default, main sequence stars track a minimal core mass that is equal to the expected core mass of a newly formed Hertzsprung gap star with mass equal to the pre-mass-transfer donor mass, scaled by the fraction of the donor's main sequence lifetime \citep{RomeroShaw:2023}, similar to the model of \citet{Neijssel:2020CygX1}.  At the end of the main sequence, the core mass is set to the greater of the currently predicted core mass and its tracked minimal core mass, not to exceed the total stellar mass.  The \optionname{ZERO} variant follows the previously used \citet{Hurley:2002} behaviour.  Meanwhile, the newly added \optionname{BRCEK} prescription applies to all forms of main-sequence mass loss, through winds as well as mass transfer.  This prescription tracks the core masses and luminosities of mass-losing main-sequence stars following the fits of \citet{Shikauchi:2024} to detailed stellar-evolution models with additional modifications to allow for a smooth transition from main sequence to Hertzsprung gap models and includes a treatment of core masses for main sequence accretors as well as donors; these modifications are described in detail in a separate publication (Br\v{c}ek et al., in prep.).

\subsection{White Dwarf accretors} 
Accretion onto a white dwarf (WD) now follows the recipes summarised in this section.  The treatment of WD accretion depends on the composition of the accreted material (hydrogen-rich or helium rich --- the latter includes naked helium stars and helium WD donors), the accreting WD (helium WD, carbon-oxygen WD, oxygen-neon WD), and the accretion rate, leading to different mass accretion efficiencies $\eta \equiv |\dot{M}_\mathrm{accretor}|/|\dot{M}_\mathrm{donor}|$.

For accreting helium WDs, we follow the StarTrack \citep{Belczynski:2008} implementation.  Hydrogen-rich material is lost in flashes ($\eta=0$) if the mass transfer rate from the donor is less than or equal to $\dot{M}_{\rm crit1}$, given in Eq.~(60) of \citet{Belczynski:2008}.  For higher mass accretion rates, we assume accumulation and complete material retention ($\eta=1$), leading to a common-envelope episode for giant donors and a merger for non-giant donors \citep{Belczynski:2008}.  Meanwhile, the accretion of helium-rich material always has $\eta=1$.  In this case, if the mass accretion rate exceeds $\dot{M}_{\rm crit2} = 2 \times 10^{-8}$ M$_\odot$ yr$^{-1}$ (defined in Section 5.7.1 of \citealt{Belczynski:2008}), we assume that the accreted material ignites in a helium flash once the conditions on the mass accretion rate specified in Eq.~(61) of \citet{Belczynski:2008} are fulfilled and the total WD mass exceeds $0.35$ M$_\odot$; this lifts the degeneracy and allows the WD to evolve as a helium main sequence star.  On the other hand, if the mass accretion rate is below $\dot{M}_{\rm crit2}$, a type Ia-like supernova occurs once the WD's mass reaches the sub-Chandrasekhar threshold given in Eq.~(62) of \citet{Belczynski:2008}.

For accreting carbon-oxygen WDs we follow the prescription of \citet{claeys2014theoretical} Appendix B, where $\eta = \eta_{\rm He}$ when accreting helium-rich material, and $\eta = \eta_{\rm H}\eta_{\rm He}$ when the accreted material is hydrogen-rich instead. However, we use fits from \citet{nomoto_thermal_2007} and \citet{Piersanti2014} when computing the critical mass accretion rates that define different accretion regimes for $\eta_{\rm H}$ and $\eta_{\rm He}$, respectively. The results presented in \citet{nomoto_thermal_2007} classify accretion regimes according to boundaries presented in their Table~5, to which we fit quadratic polynomials as follows:
\begin{gather}
    \log(\dot{M}_{\rm H,RG}~\mathrm{yr}/M_\odot) = -8.3302 + 2.8825\frac{M}{M_\odot} - 0.9802\left(\frac{M}{M_\odot}\right)^2,\\
    \log(\dot{M}_{\rm H,ST}~\mathrm{yr}/M_\odot) = -9.2176 + 3.5732\frac{M}{M_\odot} - 1.2138\left(\frac{M}{M_\odot}\right)^2,
\end{gather}
where $M$ is the WD mass.  Then $\eta_{\rm H} = \dot{M}_{\rm H,RG} / \dot{M}_\mathrm{donor}$ if $\dot{M}_\mathrm{donor} \geq \dot{M}_{\rm H,RG}$ (optically thick hydrogen winds regime);  $\eta_{\rm H} = 0$ if $\dot{M}_\mathrm{donor} < \dot{M}_{\rm H,ST}$ (hydrogen flashes regime);  and $\eta_{\rm H} = 1$ otherwise (stable hydrogen burning regime).

Meanwhile, accretion of helium-rich material onto a carbon-oxygen WD results in accretion regimes defined by Eq.~(A1) and the corresponding coefficients in Table A1 of \citet{Piersanti2014}, with the caveat that we merge their mild and strong flashes into a single flashes regime. Thus $\eta_{\rm He} = \dot{M}_{\rm He,RG/SS} / \dot{M}_\mathrm{donor}$ if $\dot{M}_\mathrm{donor}  \geq \dot{M}_{\rm He,RG/SS}$ (He-rich material accumulates and the accretor enters a giant-like regime);  $\eta_{\rm He} = 1$ if $\dot{M}_{\rm He,RG/SS} >  \dot{M}_\mathrm{donor}  \geq \dot{M}_{\rm He,SS/MF}$ (stable helium burning); $\eta_{\rm He} = 1$ if $\dot{M}_\mathrm{donor} < \dot{M}_{\rm He,SF/Dt}$ (helium accumulation); otherwise, $\eta_{\rm He}$ is given by Appendix A3 of \citet{Piersanti2014} (helium flashes), implemented as a piecewise function in COMPAS. In the helium accumulation regime, the helium shell is assumed to detonate if the WD mass exceeds 0.9 M$_\odot$ and the helium shell mass exceeds 0.05 M$_\odot$ \citep{Ruiter:2014}, leading to a supernova explosion.  Meanwhile, the stable helium burning regime leads to off-center carbon ignition if the WD mass exceeds 1.33 M$_\odot$ and $\dot{M}_\mathrm{donor} > 2.05 \times 10^{-6}$ M$_\odot$ yr$^{-1}$ \citep{Wang:2017}, forming an oxygen-neon WD.

Oxygen-neon WDs accrete according to the same prescriptions as carbon-oxygen WDs.  If the mass of an oxygen-neon WD exceeds the Chandrasekhar mass, 1.44 M$_\odot$,  it experiences accretion-induced collapse into a neutron star.

Since mass transfer from WDs allows for the mass of a WD to fall below $0.1$ M$_\odot$, we also changed the WD mass--radius relation to follow Eggleton's relation as given in Eq.~(24) of \citet{Marsh:2004}, avoiding artificially large radii at low masses while retaining the behavior of Eq.~(91) of \citet{Hurley:2000} at larger masses.


\section{Common envelopes}\label{sec:CE}

The default \COMPAS threshold for the onset of dynamically unstable mass transfer relies on the comparison of $\zeta_\mathrm{RL}$ with $\zeta_\mathrm{ad}$, the adiabatic response of the stellar radius to mass change.   We also allow dynamical instability to be decided based on one of several prescriptions for the critical mass ratio between the accretor and the donor if the \texttt{-{}-critical-mass-ratio-prescription} option is specified.  The mass transfer is labeled dynamically unstable, leading to common-envelope evolution, if the ratio of the accretor mass to the donor mass at the onset of the mass transfer episode is lower than the critical threshold.  We added the \optionname{CLAEYS} critical mass ratios following \citet{claeys2014theoretical} and the \optionname{HURLEY\_HJELLMING\_WEBBINK} critical mass ratios following \citet{Hurley:2002}.  Meanwhile, the updated \optionname{GE} and \optionname{GE\_IC} prescriptions implement the critical mass ratio models of \citet{Ge:2020a}, for the full adiabatic response and under the assumption of artificially isentropic envelopes, respectively.  These critical mass ratios are interpolated over stellar mass, metallicity, and radius (a proxy for the evolutionary stage of the star).  We also interpolate between \citet{Ge:2020a} critical mass ratios for fully conservative and fully non-conservative mass transfer, making it possible to obtain a critical mass ratio for arbitrary mass transfer efficiency, albeit under the assumption that ejected material carries the specific angular momentum of the accretor. The \optionname{GE} and \optionname{GE\_IC} prescriptions are also implemented for He-rich donors, albeit only at solar metallicity and fully conservative mass transfer.   All critical mass ratio prescriptions revert to the \citet{Hurley:2002} value of $1.59$ for WD donors.

By default, only donors with a convective envelope can survive a common-envelope episode.  However, radiative-envelope donors can now be allowed to survive if the \texttt{-{}-common-envelope-allow-radiative-envelope-survive} option is enabled.  We added a new method for determining whether a donor has a radiative or convective envelope, which can be optionally selected with the \optionname{CONVECTIVE\_MASS\_FRACTION} argument  to the \texttt{-{}-envelope-state-prescription} option.  With this choice, a donor's envelope is convective when the mass fraction of the convective outer layer (see section \ref{sec:SE}) relative to the total envelope mass exceeds a threshold set with the \texttt{-{}-convective-envelope-mass-threshold} option (default $0.1$).

The default treatment of common-envelope evolution in \COMPAS equates the energy required to unbind the envelope with the change in orbital energy, up to an efficiency parameter $\alpha$ \citep[][see Paper I]{Webbink:1984}.  The binding energy is parametrised as $G M M_\mathrm{env} / \lambda R$ \citep{deKool:1990}, where $M_\mathrm{env}$ is the envelope mass and $R$ is the total radius.  The default prescription for $\lambda$ is \optionname{LAMBDA\_NANJING}, based on \citet{XuLi:2010}, as implemented by \citet{Dominik:2012}.  We have enhanced this prescription to perform a flat extrapolation beyond the radial range where they are calibrated (necessary because \COMPAS evolutionary tracks do not perfectly match the \citealt{XuLi:2010} tracks) as well as to interpolate in mass and metallicity.  Mass interpolation is linear between mass values available in \citet{XuLi:2010} while metallicity interpolation is linear in $\log{Z}$ between their population I ($Z=0.02$) and population II ($Z=0.001$) metallicities, with flat extrapolation outside the mass and/or metallicity range.  These extrapolations and interpolations are on by default, but can be turned off by setting to false the options \texttt{-{}-common-envelope-lambda-nanjing-enhanced}, \texttt{-{}-common-envelope-lambda-nanjing-interpolate-in-mass}, and \texttt{-{}-common-envelope-lambda-nanjing-interpolate-in-metallicity}, respectively.  We have also added the option of using the effective initial mass $M_0$ \citep{Hurley:2000}, rather than the current mass, to determine $\lambda$; this can be engaged with the \texttt{-{}-common-envelope-lambda-nanjing-use-rejuvenated-mass} option.

We implemented a new, 2-stage treatment of common envelopes proposed by \citet{HiraiMandel:2022}.  Only the outer convective layer of the envelope, whose mass and binding energy are estimated following \citet{Picker:2024} (see section \ref{sec:SE}), is removed adiabatically in the first stage, using the user-specified $\alpha$ value.  Because \citet{Picker:2024} models only apply for stars more massive than $8$ M$_\odot$, we assume that for stars with mass below 2 M$_\odot$ the entire envelope is removed in the first stage, linearly interpolating the convective envelope mass for donors between 2 and 8 M$_\odot$.  The remaining portion of the envelope is assumed to be removed on the thermal timescale in the second stage, and therefore follows the angular-momentum-conserving prescription for thermal-timescale, non-conservative mass transfer, although we allow the efficiency of accretion and the specific angular momentum carried away during this second stage to be adjusted with the \texttt{-{}-common-envelope-second-stage-beta} and \texttt{-{}-common-envelope-second-stage-gamma-prescription} options, respectively.  In the rare case when both stars are simultaneously in Roche lobe overflow, the primary's radiative layer is transferred first during the second stage.  This treatment can be selected with the \optionname{TWO\_STAGE} argument to the \texttt{-{}-common-envelope-formalism} option.  

We assume no mass accretion onto a companion during a common-envelope phase by default.  However, a variety of accretion prescriptions for compact-object accretors can be chosen via \texttt{-{}-common-envelope-mass-accretion-prescription}.  A new option, \optionname{CHEVALIER}, follows model 2 of \citet{vanSon:2020} in allowing the accretor mass to grow by the significant amount $\Delta M = M_1 M_2 / (2 (M_1+M_2))$.

If either companion is in Roche lobe overflow immediately at the end of a common-envelope phase, the binary is considered to have merged during this phase.  Such binaries can now be allowed to survive if the \texttt{-{}-common-envelope-allow-immediate-RLOF-post-CE-survive} option is enabled (off by default).

Dynamically unstable mass transfer from a main sequence donor inevitably results in a binary merger.  We previously stopped evolution on a merger, but now allow the merger product of two main sequence stars only to be evolved further if the \texttt{-{}-evolve-main-sequence-mergers} option is enabled.  The fraction of total mass lost during the merger is $0.3 q / (1 + q)^2$, where $q = \min(M_1/M_2, M_2/M_1)$ \citep{Wang:2022}.  We determine the fractional main sequence age $\tau_\mathrm{frac}$ of the merger product by the fraction of hydrogen in the core, where, for simplicity, we assume that hydrogen is depleted at a uniform rate over the course of the main sequence and that the merger product is uniformly mixed.

\section{Tides}\label{sec:tides}

We added a new option to define the treatment of stellar tides, \texttt{-{}-tides-prescription}.  No tides operate in the default mode, \optionname{NONE}, except for the special case of stars that satisfy the conditions for chemically homogeneous evolution \citep{Riley:2020}: for binaries containing one or two CHE stars, the binary is circularised and the stellar rotations are synchronised to the orbital period while conserving total angular momentum.  In the \optionname{PERFECT} tides mode, the binary is re-circularised and stellar rotations are re-synchronised to the orbital period at every step of the evolution while maintaining angular momentum conservation; if no root for the new angular frequency can be found, the binary is assumed to enter a common envelope (\citealt{Darwin:1879} instability). This option applies to all stellar types, regardless of structure or compactness. Finally, our most realistic tidal interaction prescription, \optionname{KAPIL2025}, evolves the binary's semi-major axis and eccentricity and the two stellar rotation frequencies under the influence of both equilibrium and dynamical tides as described in detail by Kapil et al.~(in prep.). The option implements the orbital evolution equations from \citet{zahn_tidal_1977} as
\begin{equation}
    \frac{da}{dt} = -\frac{3}{\omega_{\rm orb}} \left(\frac{M_* + M_2}{M_*}\right) \frac{G M_2}{R_*^2} \left(\frac{R_*}{a}\right)^7 \left[ \text{Im}[k_{2,2}^2] + e^2 \left( \frac{3}{4} \text{Im}[k_{2,1}^0] + \frac{1}{8} \text{Im}[k_{2,1}^2] -5 \text{Im}[k_{2,2}^2] + \frac{147}{8} \text{Im}[k_{2,3}^2] \right) + \mathcal{O}(e^4) \right ],
\end{equation}
\begin{equation}
    \frac{de}{dt} = - \frac{3}{4} \frac{e}{\omega_{\rm orb}} \left(\frac{M_* + M_2}{M_*}\right) \frac{G M_2}{R_*^3} \left(\frac{R_*}{a}\right)^8 \left[\frac{3}{2} \text{Im}[k_{2,1}^0] - \frac{1}{4} \text{Im}[k_{2,1}^2] - \text{Im}[k_{2,2}^2] + \frac{49}{4} \text{Im}[k_{2,3}^2] + \mathcal{O}(e^2) \right],
\end{equation}
\begin{equation}
    I\frac{d\Omega_{\rm spin}}{dt} = \frac{3}{2} \frac{G M_2^2}{R_*} \left(\frac{R_*}{a}\right)^6 \left[\text{Im}[k_{2,2}^2]  + e^2 \left(\frac{1}{4} \text{Im}[k_{2,1}^2]  - 5 \text{Im}[k_{2,2}^2] + \frac{49}{4} \text{Im}[k_{2,3}^2]\right) + \mathcal{O}(e^4) \right],
\end{equation}
where $M_*$, $R_*$, $I$, and $\Omega_{\rm spin}$ are the mass, radius, moment of inertia, and rotational frequency of a given binary component, $M_2$ is the mass of its companion, $a$ is the orbital semi-major axis, $e$ is the orbital eccentricity, and $\omega_{\rm orb}$ is the orbital angular frequency. $\text{Im}[k_{l,n}^m]$ is the imaginary part of the tidal potential for a given star in the binary, and is evaluated based on the stellar type and the companion object; here, $l$ is the degree and $m$ is azimuthal wavenumber in the spherical harmonic decomposition of the tidal potential, while $n$ is the multiple of the orbital frequency in the tidal frequency. The implementation in \COMPAS enforces that tides always drive a star toward synchronisation by ignoring $\mathcal{O}(e^2)$ terms if they would increase the stellar spin past the orbital frequency. 


\section{Gravitational waves}\label{sec:GW}

Orbital evolution did not account for energy loss in gravitational waves in the \COMPAS code as described in Paper I.  Instead, evolution was stopped once two compact objects formed, ignoring the typically insignificant impact of gravitational waves emitted in wider binaries prior to compact object formation.  This remains the default behaviour, though we did change the evaluation of the time for the binary to merge through gravitational-wave radiation reaction to use the fit of \citet{Mandel:2021} (Eq.~(5)) to the numerical solution of the \citet{Peters:1964} equation (5.14).

We now allow binaries consisting of two WDs to continue their evolution with the \texttt{-{}-evolve-double-white-dwarfs} option.  To correctly evolve these and other compact binaries, we implemented gravitational-wave radiation reaction, following \citet{Peters:1964}, directly in the \COMPAS code.  The evolution of the orbital semi-major axis and eccentricity through gravitational-wave emission is enabled with the \texttt{-{}-emit-gravitational-radiation} option.


\section{General code structure}\label{sec:code}

We made a number of improvements to the overall structure of the code.  In this section, we describe the changes that enhanced input and output functionality and improved code accuracy.

\subsection{Accuracy}

We changed the default code time step durations to improve result convergence without sacrificing computational speed.  We now cap the time steps to ensure that both components in binary stars, or single stars when evolving in the SSE mode, change by no more than a fraction of 0.001 in mass due to winds or 0.1 in radius due to stellar evolution during one time step.  These default fractions can be adjusted via the \texttt{-{}-mass-change-fraction} and \texttt{-{}-radial-change-fraction} options, respectively.  We further limit the time step so that gravitational radiation reaction (see section \ref{sec:GW}) and tides (see section \ref{sec:tides}) do not change the orbit's semi-major axis by more than a fraction of 0.01 per time step.  In the case of tides, this threshold also applies to eccentricity changes and changes to the component spin frequencies.  We further reduce the time step for binaries approaching or entering Roche lobe overflow.  The time steps thresholds described here are approximate and may sometimes be exceeded by small amounts, as we estimate the time step before evolving the star and binary properties.  

Time steps can be further adjusted from the default code choices with either a constant multiplier via the \texttt{-{}-timestep-multiplier} option or more granular, phase-dependent, time step multipliers via \texttt{-{}-timestep-multipliers}; both choices are useful for debugging and detailed plotting.  The user can provide a file containing a list of desired timesteps via the \texttt{timesteps-filename} option.  We now quantize the time steps in units of $10^{-6}$ yr to improve consistency between binary and single stellar evolution.  

Integrators for quantities that require more accurate evolution within a time step, such as the orbital change on mass transfer, have been upgraded from fixed-step, first-order integration to adaptive-step, higher-order differential equation solvers from the boost library.

More coherent and robust error handling was implemented.  Improved debugging functionality and gradual option deprecation were introduced.  We changed the compiler standard from \texttt{c++11} to \texttt{c++17} and included checks for necessary libraries.

\subsection{Inputs}


The grid file functionality, which allows a user to specify the initial properties of single or binary stars to simulate rather than relying on a \COMPAS sampler or providing the initial conditions via the command line, has been augmented to allow the user to select a range of lines from a grid file with the \texttt{-{}-grid-start-line} and \texttt{-{}-grid-lines-to-process} options.

\subsection{Outputs}

All standard log files now have a record type included.  Record types make it possible to specify whether a given record is, say, a fully self-consistent record at the end of a time step or a partial record in the middle of a time step used for debugging purposes.  Users can additionally annotate log files with new program options \texttt{-{}-notes-hdrs} and \texttt{-{}-notes}.

We added the option to log high-mass X-ray binaries (HMXBs) when the \texttt{-{}-hmxr-binaries} option is enabled.  HMXBs are defined as systems with a compact object and a stellar companion that is at least 80\% Roche lobe filling, following \citet{HiraiMandel:2021}.

The logging of mass transfer tracking in the \optionname{MT\_TRACKER} record has been clarified.  The logging of additional parameters describing the strength of tidal coupling is now possible if the \optionname{KAPIL2025} prescription is used (see section \ref{sec:tides}).

A new option allows snapshots of stellar or binary properties to be logged at specified evolutionary times and/or stellar ages, as provided with \texttt{-{}-system-snapshot-time-thresholds} and \texttt{-{}-system-snapshot-age-thresholds} optional arguments, respectively.  The information to be logged in these system snapshot log files can be adjusted with the \texttt{-{}-logfile-system-snapshot-log-record-types} option.

-
-
   
\section*{Acknowledgements}
Multiple authors are supported by the Australian Research Council Centre of Excellence for Gravitational Wave Discovery (OzGrav), through project number CE230100016.
SS is a recipient of an ARC Discovery Early Career Research Award (DE220100241).  AV gratefully acknowledges support by the Marsden Fund Council grant MFP-UOA2131 from New Zealand Government funding, managed by the Royal Society Te Apārangi. IRS acknowledges the support of the Herchel Smith Fund and the Science and Technology Facilities Council grant number ST/Y001990/1. This research has made use of NASA’s \href{http://adsabs.harvard.edu/}{Astrophysics Data System Bibliographic Services}\footnote{\url{http://adsabs.harvard.edu/}}. This work used the OzSTAR and Ngarrgu Tindebeek high-performance computers at Swinburne University of Technology. OzSTAR is funded by Swinburne University of Technology and the National Collaborative Research Infrastructure Strategy (NCRIS). Ngarrgu Tindebeek is funded by the Victorian Higher Education State Investment Fund, the National Collaborative Research Infrastructure Strategy (NCRIS) through Astronomy Australia Limited (AAL), and Swinburne University of Technology. 
   
\software{COMPAS is written in C++ and  we acknowledge the use of the GNU C++ compiler, GNU scientific library (gsl), the BOOST C++ library, and the HDF5 C++ library from \url{http://www.gnu.org/software/gsl/} \citep{galassi2002gnu}. The \COMPAS suite makes use of Python from the Python Software Foundation. Python Language Reference Available at \url{http://www.python.org} \citep{CS-R9526}. In addition, the \COMPAS suite makes use of the python packages \href{http://www.astropy.org}{Astropy} \citep{2013A&A...558A..33A,2018AJ....156..123A}, \href{https://docs.h5py.org/en/stable/}{hdf5}\footnote{\url{https://docs.h5py.org/en/stable/}} \citep{collette_python_hdf5_2014}, the 
\href{http://ipython.org}{IPython}\footnote{\url{http://ipython.org}} and \href{https://jupyter.org/}{Jupyter notebook package}\footnote{\url{https://jupyter.org/}} \citep{PER-GRA:2007,kluyver2016jupyter}, \href{http://www.matplotlib.org}{Matplotlib}\footnote{\url{http://www.matplotlib.org}}  \citep{2007CSE.....9...90H},  \href{http://www.NumPy.org/}{NumPy}\footnote{\url{http://www.NumPy.org/}} \citep{harris2020array}, \href{https://www.scipy.org}{SciPy}\footnote{\url{https://www.scipy.org}} \citep{2020SciPy-NMeth}, Seaborn \citep{waskom2020seaborn}.
The \COMPAS post-processing code for detection probability currently makes use of precomputed results from the LALSuite toolkit \citep{lalsuite}, such as the {\sc{IMRPhenomPv2}} waveform \citep{2014PhRvL.113o1101H,2016PhRvD..93d4006H,2016PhRvD..93d4007K}. }

\section*{Data availability}
The living \COMPAS code is publicly available at \url{https://github.com/TeamCOMPAS/COMPAS}.   The version of record for this manuscript, \COMPAS \vCOMPAS, is released via Zenodo as \url{https://zenodo.org/records/16272773}.  We encourage the community to make results obtained with \COMPAS publicly available at  \url{https://zenodo.org/communities/compas/}. 

\bibliographystyle{aasjournal}
\bibliography{Mandel, bib}

\end{document}